# Interdependent linear complexion structure and dislocation mechanics in Fe-Ni

Vladyslav Turlo [1,2], Timothy J. Rupert [2,3,*]

[1] Laboratory for Advanced Materials Processing, Empa, Thun, 3603, Switzerland
[2] Department of Mechanical and Aerospace Engineering, University of California, Irvine, CA 92697, USA
[3] Department of Materials Science and Engineering, University of California, Irvine, CA 92697, USA
* Email: trupert@uci.edu

**Abstract:** Using large-scale atomistic simulations, dislocation mechanics in the presence of linear complexions are investigated in an Fe-Ni alloy, where the complexions appear as nanoparticle arrays along edge dislocation lines. When mechanical shear stress is applied to drive dislocation motion, a strong pinning effect is observed where the defects are restricted by their own linear complexion structures. This pinning effect becomes weaker after the first dislocation break-away event, leading to a stress-strain curve with a profound initial yield point, similar to the static strain ageing behavior observed experimentally for Fe-Mn alloys with the same type of linear complexions. The existence of such a response can be explained by local diffusion-less and lattice distortive transformations corresponding to $L1_0$-to-B2 phase transitions within the linear complexion nanoparticles. As such, an interdependence between linear complexion structure and dislocation mechanics is found.

## 1. Introduction

Linear complexions are thermodynamically-stable nanoscale phases recently discovered at dislocations in the Fe-9 at.% Mn alloy [1]. Similar to interfacial complexions confined to grain boundary regions [2–4], linear complexions are defined by a structure and chemistry that are different from the matrix yet can only exist in the presence of crystalline defects, with dislocations serving that role for linear complexions. Using atomistic simulations, the authors of this work recently predicted a wide variety of linear complexions in body centered cubic (BCC) and face centered cubic (FCC) metals [5–7]. One interesting feature of linear complexions in BCC Fe-based alloys is the presence of a metastable phase in the dislocation segregation zone, which maintain coherent interfaces with the matrix phase. These metastable phases have been reported for the Fe-Ni system with simulations [5] and for the Fe-Mn system with experiments [8]. Other interesting features have been predicted for FCC alloys, such as the formation of 2D platelet phases which can form platelet arrays along partial dislocations or replace the dislocation stacking fault [7]. While some of these complexion types still require experimental validation, it is clear that linear complexions at dislocations represent an new exciting materials research area for crystalline solids, as this topic has the potential to enable new materials with unique properties.

While the effects of grain boundary complexions on various material properties have been studied extensively [9–12], similar research on the influence of linear complexions is limited to the work of Kwiatkovski de Silva et al. [13], who demonstrated a static strain aging effect in single crystal Fe-Mn samples containing linear complexions. Specifically, the atomic-scale details of dislocation-linear complexion interactions and the associated mechanical behavior are not known. Atomistic simulations are proven to be a powerful tool for investigating the nanoscale mechanics involving dislocation interactions with alloying elements [14], grain boundaries and grain boundary complexions [15,16], nanoscale precipitates [17,18], ceramic nanoparticles [19], Guinier-Preston zones [20,21], and vacancy clusters [22]. Atomistic simulations act in these situations as a digital microscope, providing a great level of detail on structural and chemical transitions as well as deformation mechanisms at the nanoscale. For example, multi-principal element alloys have intriguing mechanical properties yet their deformation physics are complicated by the compositional complexity of the lattice. Jian et al. [23] reported on the roles of lattice distortion and chemical short range order on dislocation behavior, finding that these factors can result in enhances glide resistance. Xu et al. [24] explored the local slip resistances in a BCC multi-principal element alloy on a variety of slip planes and with a variety of Burgers vectors, observing that these alloys could deform by a multiplicity of slip modes. The work of Wang et al. [14] provided experimental validation of such a plasticity mechanism and connected this behavior to the observation of a strength plateau at intermediate temperatures, rather than the rapidly decreasing strength of traditional BCC alloys with increasing testing temperature.

Due to the recent discovery of linear complexions, a comprehensive investigation of their effect on dislocation propagation and pinning is missing in the literature. In this paper, we provide the first mechanistic insight into the effect of nanoparticle array linear complexions in a BCC Fe-Ni alloy on mechanical behavior, with a solid solution of the same composition providing a point for comparison. The atomistic mechanisms associated with dislocation pinning and unpinning events during the shear deformation are investigated in detail, and connected to the shape of the stress-strain curve. Finally, the structure of the linear complexion is found to change as the dislocation and its local stress field moves away, resulting in an interdependence of dislocation behavior and complexion structure. The results shown here highlight that linear complexions are defect states that both alter and react to the dislocation environment, providing a pathway for the direct manipulation of mechanical behavior.

## 2. Materials and Methods

Atomistic simulations, including molecular statics, molecular dynamics (MD), and hybrid Monte Carlo (MC)/MD, were performed using the Large-scale Atomic/Molecular Massively Parallel Simulator (LAMMPS) software [25], with an embedded-atom method (EAM) potential parametrized to reproduce the binary Fe-Ni system used to model atomic interactions [26]. All MD simulations used a 1 fs integration timestep. Atomic snapshots were analyzed and visualized using the OVITO software [27]. Crystalline structure and chemical ordering were analyzed using the Polyhedral Template Matching method [28], while the positions of dislocation lines were identified using the Dislocation Extraction Algorithm [29].

The Fe-Ni system and the given potential were chosen for several reasons. First, the Fe-Ni phase diagram is similar to the phase diagram of the Fe-Mn system, in which linear complexions were first discovered experimentally. This is important because only two Fe-Mn potentials currently can be found in the literature, but neither is appropriate for the goals of this study. Bonny et al. [30] developed an Fe-Mn EAM potential while Kim et al. [31] developed an Fe-Mn modified embedded atom method (MEAM) potential but neither was rigorously fitted to reproduce the bulk phase diagram. Moreover, MEAM potentials are not currently compatible with the hybrid MC/MD code used here. Since it is critical to reproduce the relevant phases for the alloy system, the Fe-Ni potential from Bonny et al. [26] was chosen for this work as it was fitted based on the experimental phase diagram for Fe-Ni and was found to reproduce the stable intermetallic phases ($L1_0$-FeNi and $L1_2$-FeNi$_3$). The main weakness of the potential is that the solubility limit of Ni in BCC Fe is overestimated, meaning that exact composition values for complexion transitions should not be compared with experiments. In addition, Domain and Becquart [32] performed density functional theory (DFT) calculations of segregation to self-interstitial atom clusters, finding that Ni may be able to segregate to sites under both local compressive and tensile stresses, while the EAM potential only predicted segregation to the tensile regions. This suggests that some segregation of Ni dopants to both sides of the dislocation may occur during the initial segregation stages. However, the final linear complexion structure is not expected to be affected by this, and in fact the simulated linear complexions resemble those observed experimentally in Fe-Mn [5,6]. Strong variations of Ni composition along the dislocation line were observed at the compression side in the dislocation segregation zone, with the composition near precipitates approaching ~50 at.% Ni while in between precipitates the composition was near the global composition in the system. Based on these observations, we conclude that the final form of the complexion and solute distribution around the dislocation core is controlled by the second-phase precipitation and growth, and not by the initial solute segregation. Next, we provide the simulation details for a representative computational cell containing linear complexions, prepared for mechanical testing.

First, an initial simulation cell with two edge dislocations was prepared, as shown in Figure 1(a). The dislocations were inserted by removing one-half of the atomic plane in the middle of the sample and relaxing the atomic structure using the conjugate gradient descent method implemented in LAMMPS. Next, 2 at.% Ni atoms were randomly distributed within the sample by replacing Fe atoms, followed by MD equilibration with an NPT ensemble (constant number of atoms, constant pressure, and constant temperature) at zero pressure and 300 K temperature for 20 ps. Three thermodynamically-equivalent initial solid solution configurations with different random distributions of solutes were prepared with this procedure, providing a baseline for comparison against samples containing linear complexions. To induce linear complexion formation, the alloy samples were equilibrated at 500 K using the hybrid MC/MD method which allows for both chemical segregation as well as local structural relaxations. The MC steps were performed after every 0.1 ps of MD relaxation time, using a variance-constrained semi-grand canonical ensemble that can stabilize alloy systems with coexisting phases [33,34]. This method has been used in a number of recent modeling studies to capture complexion transitions in metallic alloys [35–37]. The MC/MD procedure led to Ni segregation to the compressive side of the dislocations and then formation of linear complexions in the form of nanoscale precipitate arrays composed of metastable B2-FeNi and stable $L1_0$-FeNi phases, as shown in Figure 1. The presence of nanoscale precipitates reduces the compressive stresses on the side of the dislocation with the extra half-plane of atoms (see the zoomed view of the bottom dislocation in Figure 2). More details about the MC/MD procedure for linear complexions in the Fe-Ni system can be found in our previous studies of equilibrium complexion states [38,39]. The MC/MD simulations were determined to be in equilibrium once the rate of evolution of total simulation cell energy fell below 1 eV/ns, although the procedure was continued for additional time to sample three distinct yet thermodynamically equivalent samples. These samples were then cooled to 300 K over 20 ps with MD for subsequent mechanical testing.

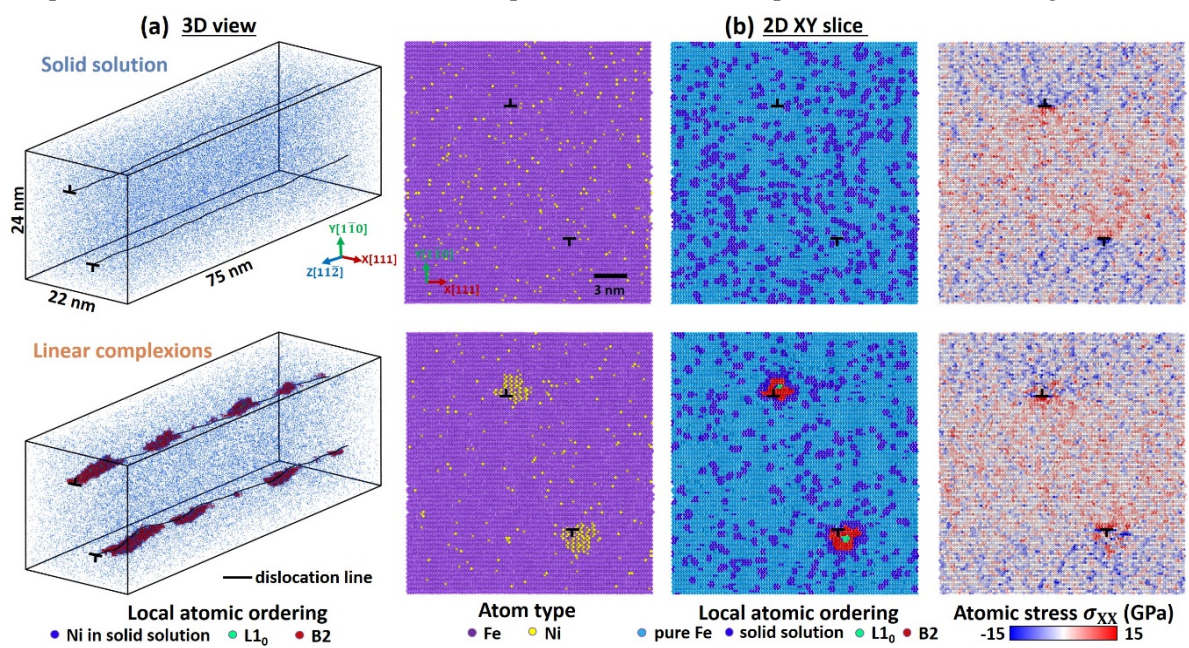

**Figure 1.** (**a**) 3D perspective view of the simulation cells with the random distribution of solutes, and with linear complexions comprised of nanoscale precipitates containing B2 (red) and $L1_0$ (green) phases. Atoms are colored by their local atomic order and the matrix Fe atoms are deleted for visualization purposes. The dislocations are shown as black lines. (**b**) XY slices of the system with Z-position of 28 nm and the thickness of 0.26 nm. The atoms are colored as follows: magenta – Fe, yellow – Ni, light blue – BCC pure Fe, dark blue – BCC Fe-Ni solid solution, red – B2-FeNi, and green – $L1_0$-

FeNi. The XX component of the atomic stresses are shown in red-white-blue color scheme with red indicating tensile stresses and blue indicating compressive stresses.

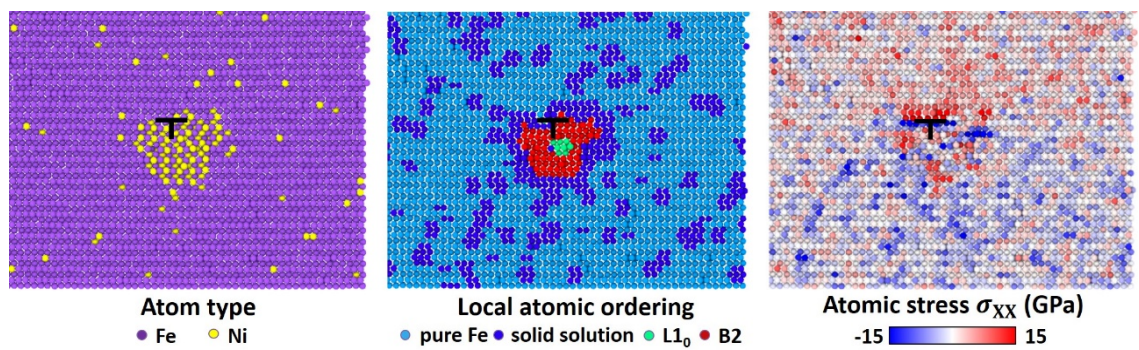

**Figure 2.** Zoomed view of the bottom dislocation in the system with linear complexions that was previously shown in Figure 1(b).

The solid solution specimens and the samples containing linear complexions were then deformed to promote dislocation slip by applying XY shear deformation to the simulation cell. The deformation of the cell was performed with the non-equilibrium MD method [40,41], using an engineering strain rate of $10^8$ s$^{-1}$ at 300 K. To test the effect of deformation rate, an additional simulation was performed at an engineering strain rate of $10^7$ s$^{-1}$. The deformation of the cell was stopped after 10% applied shear strain, and the stress-strain curves were extracted and analyzed in the context of atomic-scale deformation mechanisms and the local structure of the linear complexions.

## 3. Results and Discussion

Figure 3(a) shows the obtained stress-strain curves for the samples with linear complexions. We observe profound initial stress peaks for all the samples, indicated by black circles. The initial plastic events are followed by smaller peaks, indicated by black triangles, that represent the flow stress of the dislocations passing by the linear complexions. Similar stress-strain curves with aprofound first peak have been previously reported in experimental work on linear complexions in an Fe-9 at. % Mn alloy [13]. The substantial strain aging is observed in both experimental Fe-Mn and modeled Fe-Ni systems with linear complexions. The average values for both the initial break-away stress and the flow stress for the simulated Fe-Ni samples with linear complexions are presented in Figure 3(b). The mean initial break-away stress of 586 MPa is almost 50% higher than the mean flow stress of 404 MPa. We note that these values should not be directly compared to experimental measurements, as other aspects of alloy strengthening from defects such as grain boundaries are not present in the simulation cell, which isolates the dislocation pair.

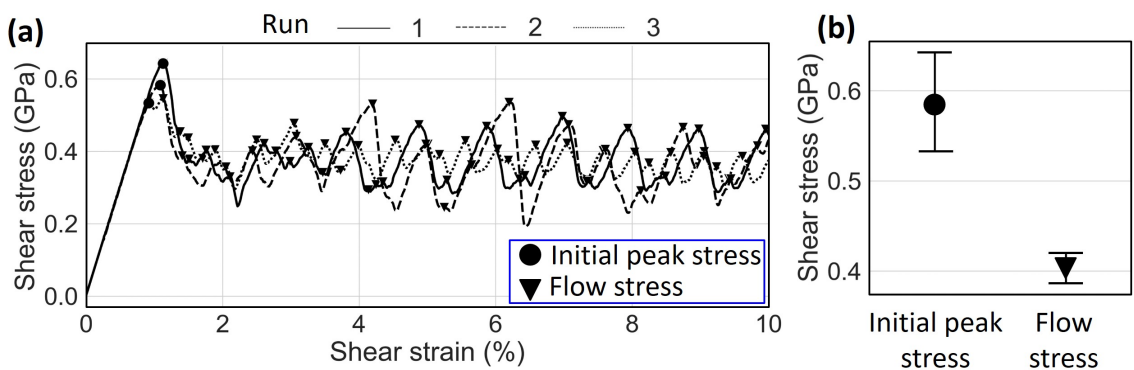

**Figure 3.** (**a**) The stress-strain curves for three samples with linear complexions. The initial peak stresses and flow stresses are denoted by circle and triangle symbols, respectively. (**b**) Mean values and corresponding standard deviations for the initial peak stress and flow stress.

To understand how deformation differs between the solid solution and linear complexion states, Figure 4 presents the initial flow events for representative examples of each of the two sample types. Figure 4(a) shows the elastic loading and initial flow event, where it is clearly observed that the dislocation can move much more easily in the solid solution sample. Figure 4(b) shows the dislocation pair at three different simulation times (or, equivalently, applied strains since the strain rate is controlled). The two dislocations in the pair remain relatively straight during the motion, with the top dislocation shifting to the right and the bottom dislocation moving to the left. The solutes in the solution act as obstacles that must only be locally overcome, leading to very little change in the dislocation shape. The bottom dislocation is shown from a top view in the lower half of Figure 4(b), demonstrating the progressive migration that leads to a temporary stress drop as a set of local, solute obstacles are overcome. The local obstacles are easily overcome, which is why the initial peak stress for the solid solution specimens is relatively low. In general, the dislocation in the solid solution sample behaves in a "textbook" fashion, with few features of interest.

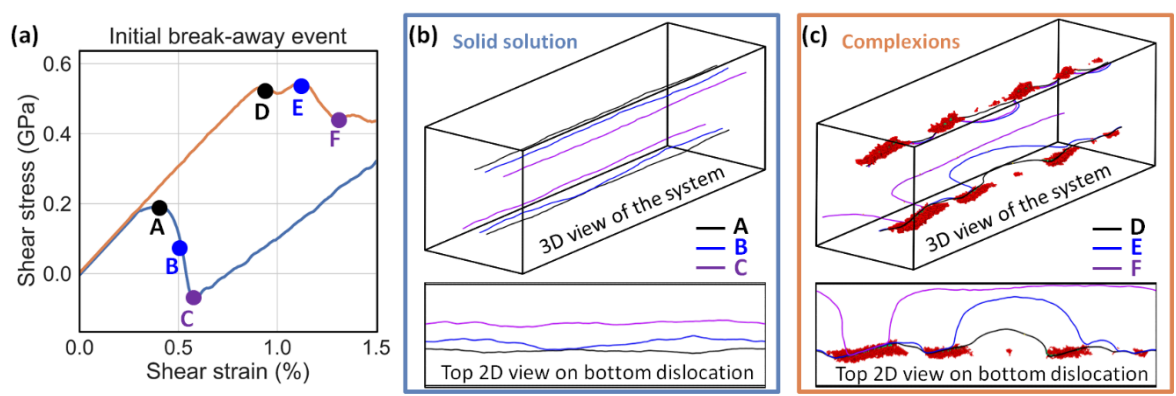

**Figure 4**. (**a**) Representative stress-strain curves showing the initial break-away event for a sample with solid solution (blue) and a sample with linear complexions (orange). 3D perspective views of the entire computational cell and 2D views of the bottom dislocation are shown for (**b**) the solid solution and (**c**) the sample with linear complexions. The position of the dislocations are shown as solid curves of different colors corresponding to different times, as shown in (**a**).

The mechanism of the dislocation motion is drastically different for the specimen with linear complexions, with dislocation bowing and unpinning from the nanoscale precipitates one by one, starting in the region with the largest distance between particles, controlling the yield event shown in Figure 4(c). It is important to note that only one dislocation, the defect at the bottom of the cell, is moving in the linear complexion sample. The other dislocation at the top of the cell bows under increasing stress but remains pinned by the nanoparticle array linear complexion. One dislocation is favored to move over the other because the nanoparticle arrays, while similar, are not exactly identical. A bowing mechanism is easiest in the location where there is the largest spacing between obstacles, which occurs near the middle of the lower dislocation. While dislocation bowing is a common mechanism for overcoming precipitates in conventional alloys, an important distinction is found for the linear complexion state: the obstacle is not in the dislocation's slip plane. Traditional Orowan bowing occurs when an inpenetrable obstacle (e.g., a precipitate with a different crystal structure than the matrix) impedes the dislocation's slip path, requiring bowing to move past the obstacle in a way that leaves dislocation loops around the obstacle. However, in the case of linear complexions, the nanoparticles are primarily above or below the dislocation slip plane (depending on whether one is looking at the top or bottom dislocation). Even the small portion of the B2-FeNi intermetallic phase that crosses the the slip plane in Figure 1 has the same BCC crystal structure as the matrix phase and a lattice parameter that is similar. The linear complexion is not an impenetrable obstacle and in fact no new dislocation loops or segments are formed as the dislocation pulls away. The strong initial pinning can be explained from the same energetic perspective that is used to describe the complexion nucleation. Segregation of Ni occurs to the compressed region near the edge dislocation, until the local composition is enriched enough that a complexion transition can occur [5,38]. Although the Gibbs free energy of the $L1_0$ phase is lower, which is why this structure appears on the bulk phase diagram, the restriction to a nanoscale region and the large energy cost for an incoherent BCC-$L1_0$ interface results in the formation of a "shell" of a metastable B2 phase surrounding the $L1_0$ core [38]. Fundamentally, the local stress field near the dislocation is relaxed by this transformation and the driving force for motion under shear stress (i.e., the Peach-Koehler force [42,43]) is therefore reduced. We do note that segregation of Ni to the dislocation is needed to create the linear complexion states, thus lowering the solute composition in the matrix solid solution and removing some weak obstacles. However, the net effect is still a notable strengthening increment, as the strong linear complexion obstacles more than make up for losing some amount of solid solution strengthening.

Since the periodic boundary conditions are used here, the dislocation exits the cell after it breaks away from the linear complexion and then reenters on the other side, where it eventually interacts with the complexon again. Therefore, subsequent dislocation-complexions can be observed by continuing the simulation and investigating the flow events at larger strains. A detailed look at multiple flow events is shown in Figure 5 for one of the linear complexion samples. In addition to the stress-strain curve shown in black, measurements of the dislocation length at any given time are also extracted and presented as the blue curve. Three separate events are labels A-B, C-D, and E-F and isolated into separate parts of the figure. Similar to the initial yield event in Figure 4, only one dislocation (bottom) is moving while the other (top) remains pinned by the nanoscale precipitates. Figure 5 shows that there are cyclic undulations in the shear stress, which correlate with the observation of repeated bursts in the dislocation density that are associated with dislocation bowing. In fact the shapes of the dislocation length bursts are extremely similar for each cycle, suggesting that the physical events are similar as well. Snapshots A-B show the final pinned segment of the dislocation breaking away from a large particle in the complexion array, then enerting from the other side and actually being attracted to the particles, as evidenced by the fact that the dislocation is pulled closer to the precipitates first in frame B. Snapshots C-D show the first bowing event in a new cycle, which occurs at the location with the spacing between particles and is reminiscent of the events presented in Figure 4(c). We do remind the reader that this bowing does occur at lower stresses than the initial yield event, perhaps suggesting that something about the complexion has evolved. Finally, snapshots E-F show an intermediate bowing event, neither the first nor the last in a cycle.

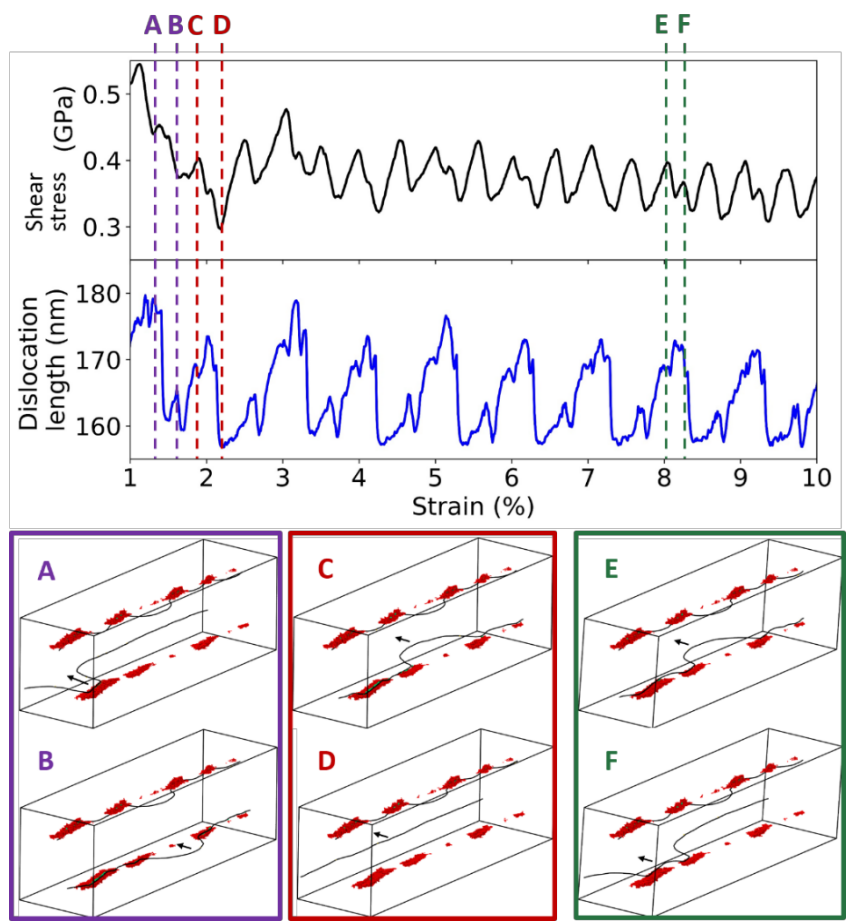

**Figure 5.** The evolution of the simulation cell shear stress (top, black) and dislocation length (bottom, blue) for the sample with linear complexions, with different atomic configurations sampled on the right side of the figure. The atomic configurations are grouped by different passes of the dislocation that happen due to the periodic boundary conditions of the simulation cell. Black arrows denote the direction of motion for mobile dislocation at the bottom of the cell.

To investigate complexion structure during the deformation simulation, the number of B2 and $L1_0$ atoms in the simulation cell were tracked and are shown in Figure 6(a). A cyclic pattern is again observed, suggesting a connection to the repetitive dislocation motion through the cell. Most notable is that reductions in $L1_0$ atoms are generally aligned with increases in the number of B2 atoms (and vice versa). Figures 6(b)-(e) show atomistic snapshots of important dislocation interactions with a linear complexion particle, in an effort to understand this cyclic behavior. In Figure 6(b), the dislocation is still pinned next to the $L1_0$ precipitate, so the complexion contains the expected $L1_0$ core and B2 shell. The dislocation started to pull away from the particle in Figure 6(c), and close inspection of the complexion particle shows a reduction in the size of the green $L1_0$ region. However, the transition to the B2 structure does not happen immediately, as some time is apparently needed, although short since it is captured on MD time scales. Figure 6(d) shows a later time when the dislocation has moved fully away from the particle, and the linear complexion is almost entirely composed of B2 structure in this frame (a very small number of green $L1_0$ atoms remain but the number is dramatically reduced). This observation provides further support for the concept that the two-phase complexion structure is caused by the dislocation's hydrostatic stress field. When that stress field is no longer present, a diffusion-less and lattice distortive transformation from an FCC-like structure ($L1_0$) to a BCC-like structure (B2) occurs. Figure 6(e) shows that the $L1_0$ region of the precipitate starts to be recovered as the dislocation, and its stress field, arrives at the other side. We note that while a cyclic behavior is observed in Figure 6(a), the number of B2 atoms trends downward generally as the process continues. This could be a sign that the linear complexion particles are becoming smaller in subsequent cycles (or at least during the transitions from the first few cycles to the steady-state cycling), which would provide an explanation for why the initial yield event is more difficult than later dislocation flow. Figure 6 generally highlights the close connection between the structure of the linear complexion and its interactions with the dislocation. The complexions restrict the dislocation and make it harder to move, while the structure of the complexion relies on the local stress field from the dislocation to find a local equilibrium structure. There is hence an interdependence of these two defect structures, which truly sets linear complexions apart from traditional obstacles to dislocation motion.

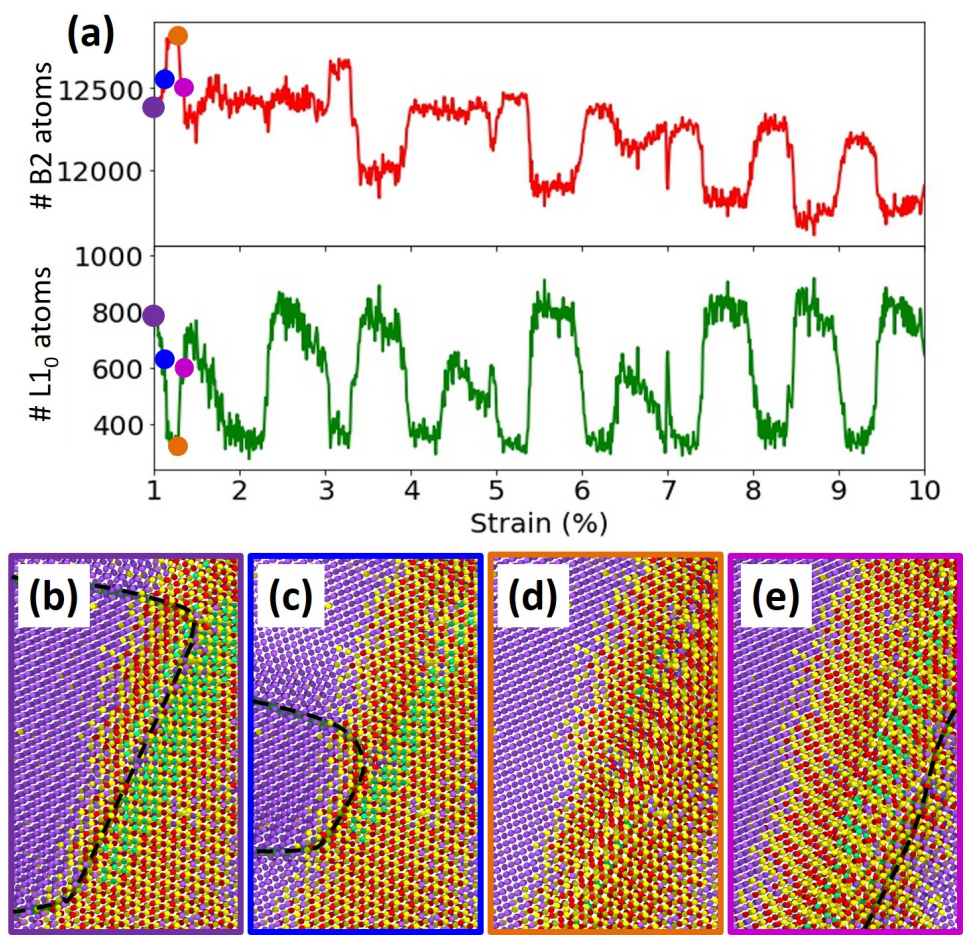

**Figure 6.** (**a**) The number of B2 (red) and $L1_0$ (green) atoms in the computational cell containing linear complexions, as the deformation simulation progresses. (**b**)-(**d**) A region perpendicular to the Y-direction from the dislocation slip plane to the middle of the precipitate. Yellow spheres represent Ni atoms, while violet, red, and green spheres represent Fe atoms in the matrix, B2, and L10 phases, respectively. The dashed line shows the position of the dislocation line. A diffusion-less and lattice distortive transformation is observed, with the complexion structure being dependent on the dislocation position.

Finally, to show that this local transition within the complexion is not dependent on the deformation rate used here, an additional deformation simulation was run at one order of magnitude slower strain rate ($10^7$ s$^{-1}$). The results of this simulation are presented in Figure 7, capturing all of the same important features described above. Loading is initially elastic until a high stress is reached, when the dislocation is able to bow out at the region with the largest spacing between linear complexion particles. The dislocation re-enters the simulation cell and becomes pinned, and subsequent dislocation unpinning and motion follows the same mechanism. A reduction of the number of $L1_0$ atoms in the system and a corresponding increase in the number of B2 atoms is observed each time the dislocation is able to break away. These trends agree with the $L1_0$-to-B2 lattice distortive transformation shown in Figure 6.

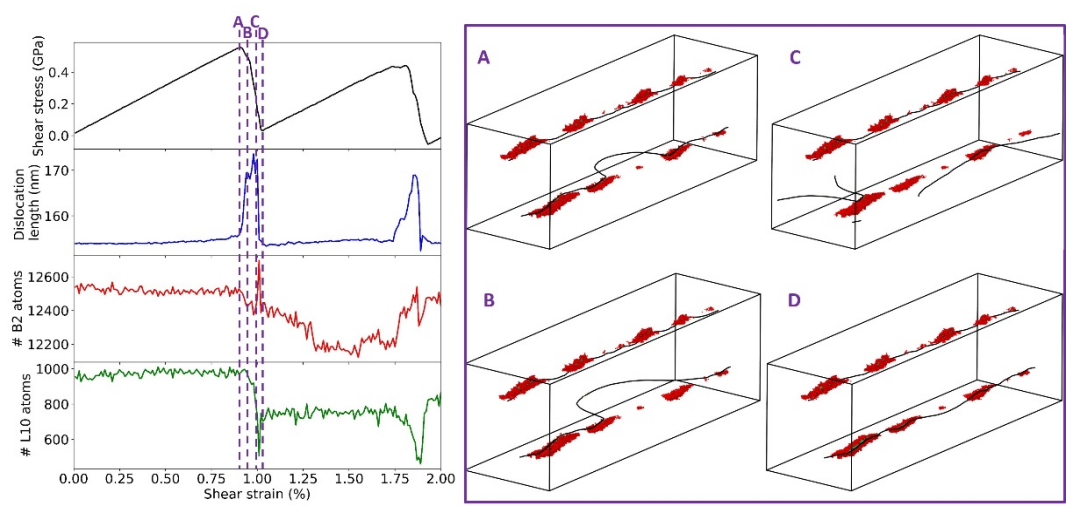

**Figure 7.** The evolution of the simulation cell shear stress (black), dislocation length (blue), the number of B2 (red) and L10 (green) atoms for the sample with linear complexions deformed at a strain rate of 10$^7$ s-1, with different atomic configurations sampled on the right side of the figure for the initial break-away event of the dislocation.

## 4. Conclusions

This paper presents the first atomistic study of the effect of linear complexions on the mechanical behavior of metallic alloys, with Fe-Ni used as a model system. A strong pinning effect of linear complexions on their host dislocations is observed, which is connected to the alteration of the dislocation's stress field in the crystal. This pinning effect leads to a substantial increase in the initial breakaway stress, with a pronounced initial peak stress is in agreement with experimental observations from alloys with similar complexion structures. Dislocation motion away from the nanoparticle arrays leads to an L1$_0$-to-B2 lattice distortive transformation as the stress field which stabilized the original complexion structure is removed. These findings provide additional understanding and context for nanoscale phase transformations induced by dislocation stress fields and their effect on the mechanical properties of materials. Additional study to obtain a complete understanding of linear complexion thermodynamics and the deformation physics associated with dislocation complexion-interactions are needed to enable "defects-by-design" which can be used to tailor mechanical response.

**Author Contributions:** Conceptualization, Methodology, Data curation, Formal analysis, and Writing-original draft, V.T.; Conceptualization, Methodology, Resources and Writing - review & editing, T.J.R. All authors have read and agreed to the published version of the manuscript.

**Funding:** This research was funded by the U.S. Army Research Office, grant number W911NF-16-1-0369.

**Conflicts of Interest:** The authors declare no conflict of interest.